\documentclass[journal]{IEEEtran}
\IEEEoverridecommandlockouts
\usepackage{cite}
\usepackage{amsmath,amssymb,amsfonts}
\usepackage{algorithmic}
\usepackage{graphicx}
\usepackage{textcomp}
\usepackage{xcolor}
\usepackage{multirow, makecell, caption}
\usepackage{tabularx}
\usepackage{adjustbox}
\usepackage{float}
\usepackage{multicol}
\usepackage{lipsum}
\usepackage{comment}
\usepackage{dblfloatfix} 
\usepackage{subcaption}
\usepackage[justification=centering]{caption}

\def\BibTeX{{\rm B\kern-.05em{\sc i\kern-.025em b}\kern-.08em
		T\kern-.1667em\lower.7ex\hbox{E}\kern-.125emX}}
\begin{document}
	\title{Diagnosis of COVID-19 disease using CT scan images and pre-trained models}

	\author{\IEEEauthorblockN{Faezeh Amouzegar, Hamid Mirvaziri, Mostafa Ghazizadeh-Ahsaee, Mahdi Shariatzadeh}\\
		\vspace*{0.5cm}
		\IEEEauthorblockA{\small{\textit{Department of Computer Engineering}, \textit{Shahid Bahonar University of Kerman}, Kerman, Iran\\
				\textit{amouzegar@eng.uk.ac.ir, hmirvaziri@uk.ac.ir, mghazizadeh@uk.ac.ir, shariat@eng.uk.ac.ir}}}}

	\IEEEabstract{Diagnosis of COVID-19 is necessary to prevent and control the disease. Deep learning methods have been considered a fast and accurate method. In this paper, by the parallel combination of three well-known pre-trained networks, we attempted to distinguish coronavirus-infected samples from healthy samples. The negative log-likelihood loss function has been used for model training. CT scan images in the SARS-CoV-2 dataset were used for diagnosis. The SARS-CoV-2 dataset contains 2482 images of lung CT scans, of which 1252 images belong to COVID-19-infected samples. The proposed model was close to 97\% accurate.}
	
	\IEEEkeyword{ResNet, GoogleNet, ShuffleNet, SARS-CoV-2 dataset, negative log-likelihood loss}
	
	\maketitle
	
	\section{Introduction}
	Coronavirus disease 2019 (COVID-19) is an infectious disease causing severe respiratory syndrome. The first patient was detected in Wuhan, China, in December 2019. The illness spread around the world, leading to the COVID-19 pandemic
	\cite{sun2020understanding}.
	The main symptoms of this illness are malaise, dyspnea, fever, cough, headache, and a reduced sense of taste and smell
	\cite{struyf2022signs}.
	Symptoms may appear as late as two weeks after the patient is infected. Currently, at least one-third of infected people have no serious symptoms.
	
	One way to diagnose COVID-19 disease is to perform a CT scan of the lungs of suspected people. Because CT scans are not available everywhere or cannot be used when the patient is in the ICU, radiographic imaging is also used to diagnose this disease
	\cite{alsharif2021effectiveness}.
	
	At present, radiographs and CT scans of the lungs are evaluated qualitatively by radiologists to decide on the presence of the COVID-19 disease and its severity or the effectiveness of treatment
	\cite{lyu2020performance}.
	
	Today, several methods, such as clinical signs associated with the disease, and more accurate diagnostic methods, such as CT scans of the lungs, are used to diagnose COVID-19
	\cite{song2021deep}. This study aimed to achieve an accurate diagnostic method for intelligent and automatic diagnosis of COVID-19 disease by using lung CT scan image processing techniques and using the results of this method as an accurate diagnostic tool.
	
	In section
	\ref{Related works}, related works are introduced. Section
	\ref{Proposed Approach} presents the data sets, loss function, and model backbones. Experimental results are presented in Section \ref{Experiments}, and conclusions are made in Section \ref{Conclusion}.

	\section{Related Works}
	\label{Related works}
	Zhao et al. \cite{wang2020contrastive}
	proposed a joint learning system for precise recognizable proof of COVID-19 using viable learning, heterogeneous datasets, and dispersed contrasts. They built a solid spine by overhauling COVIDNet in angles of the organized engineering and learning technique to make strides in forecast precision and learning productivity. They unequivocally illuminated the space exchange between locales by standardizing isolated usefulness in covered-up spaces. Besides, they recommended employing a contrast instruction crystal to extend the run of invariant semantic integration to move forward with the classification of each dataset. They created and evaluated a strategy using a combination of two large COVID-19 demonstration datasets using CT images. Their tests showed that their approach improved the execution of both datasets.
	
	Xu et al. \cite{james2021exploring}
	used a version of the convolutional neural network (CNN) to categorize people with COVID-19. They employed templates trained in advance for the system. Three hospitals in China provided the data. In this study, they examined 618 CT images. Of these, 219, 22, and 175 people had COVID-19, had influenza virus pneumonia, and were the general public, respectively. Of the samples, 85\% (528) were used to train, and the remaining samples were employed to test the resulting model. The frame results showed 86.7\%, 81.5\%, 80.8\%, and 81.1\% accuracy, sensitivity, precision, and  F-measure, respectively.
	
	Javaheri et al. have developed a method of deep learning called \cite{javaheri2020covidctnet} Covid CTNet to detect coronavirus infections through CT images. Their system used the BCDUNet architecture developed on UNet. COVID-19 is distinguished from CAP and other lung diseases in their scheme. In their experiments, the system used a total of 89,145 CT images. Their dataset was split into 90\% for training and 10\% for testing. Experimental results showed that their developed system achieved 91.66\%, 87.5\%, 94\%, and 95\% accuracy, sensitivity, specificity, and AUC, respectively.
	
	\begin{figure*}[]
		\centerline{\includegraphics[]{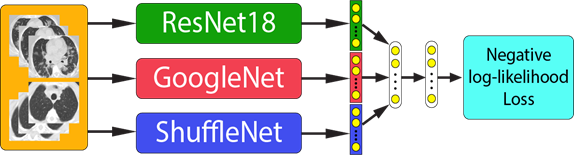}}
		\caption{The architecture of proposed method.}
		\label{fig1}
	\end{figure*}
	
	Cifci \cite{cifci2020deep} suggested another scheme for this problem using deep transfer learning. He conducted his study using CT images and used AlexNet and InceptionV4 as his pre-trained models as they are popular for analyzing medical images. They acquired 5800 CT images from the public repository to develop the system. His training set contained 4640 CT images, and his test set contained 1160 images. His experimental results showed AlexNet to perform relatively stronger than Inception V4. With a specificity of 87.45\% and sensitivity of 87.37\%; the overall accuracy of AlexNet was 94.74\%.
	
	Loey et al. \cite{loey2020within} used the Generative Adversarial Network (GAN), Resnet18, Googlenet, and Alexnet to present their system. GAN was used to generate more images and make up for the small number of COVID-19 X-ray images necessary for proper detection. The number of those images was 307, and they were divided into four classes: pneumonia\_bac, pneumonia\_vir, COVID-19, and normal.
	
	The system Horry et al. \cite{horry2020x} developed uses pre-trained model concepts in X-ray images. Their proposed system used four common pre-trained models: Inception, Xception, VGG, and Resnet. Their dataset contained 100 COVID-19, 100 pneumonia, and 200 healthy experimental samples. They used an 80:20 training and testing set ratio. Their experimental results showed that the VGG19-based model achieved 83\%, 80\%, and 80\% accuracy, sensitivity, and F-score, respectively, which was the best performance in their studies.
	
	Sethy and Behra
	\cite{sethy2020detection}
	used support vector machines (SVMs) and pre-trained CNN models to introduce a COVID-19 diagnosis system. This algorithm performed classification by automatic feature extraction using two separate datasets, SVM, and 11 pre-trained CNN models. Their dataset contained 133 images as positive samples and 133 plain radiographs as negative samples. Their experimental results showed that Resnet 50 achieved 90.76\% kappa, 91.41\% Matthews correlation coefficient (MCC), 95.52\% false-positive rate (FPR), and 95.38\% accuracy on SVMs.
	
	Minaee et al.
	\cite{minaee2020deep}
	developed a model called DeepCOVID that used deep transfer learning for predicting COVID-19 by X-ray images. In their research, four known pre-trained models, including ResNet18, ResNet50, SqueezeNet, and DenseNet121, were employed for diagnosing COVID-19. Of the 5071 open access images of their dataset, 2000 were used for training, and 3000 images were used for testing, and they used the COVIDXray5k dataset. The best result of their model was 100\% sensitivity and 95.6\% specificity using SqueezeNet.
	
	Abbas et al.
	\cite{abbas2021classification}
	distinguished COVID-19 patients from healthy cases using a system called DeTraC deep ResNet18. By examining class boundaries and using class decomposition approaches, anomalies in image datasets could be fixed by DeTraC. Their system was trained by 196 samples which were 80 cases for normal class, 105 cases for COVID-19, and 11 cases of SARS. Overall, 1764 samples were generated by their system. They allocated 70\% of their dataset to the training set and 30\% to evaluation and achieved 91.87\% specificity, 97.91\% sensitivity, and 93.36\% accuracy using the system they proposed.
	
	Moutounet-Cartan
	\cite{moutounet2020deep}
	introduced a system using X-ray images based on deep learning for COVID-19 diagnosis. Their system used famous models, such as VGG16, VGG19, InceptionResNetV2, InceptionV3, and Xception. Overall, 327 x-rays images were used. Their dataset contained 152 healthy cases, 125 COVID-19 cases, and 50 cases from other pneumonia disorders. Their dataset was split for 5-fold cross-validation to consider AUC and sensitivity for COVID-19. VGG16 was the best model, with 84.1\% accuracy, 87.7\% sensitivity, and 97.4\% AUC.
	
	The system proposed by Hemdan et al.
	\cite{hemdan2020covidx},
	called COVIDXNet, aimed to diagnose coronavirus infection based on X-ray images and CNNs. Seven pre-trained models were used in their work. Their dataset contained 50 images, half from healthy persons and half from COVID-19 patients. Their dataset was split into 80\% and 20\% parts for the training and test sets, respectively. Experimental results showed that Inception V3 achieved the worst results. DenseNet and VGG19 were the best models, with 91\% F-measure and 90\% accuracy.
	
	Modified Inception was a technique used by Wang et al.
	\cite{wang2021deep}
	to develop a model to reduce the feature dimensions prior to the final classification. They used 1040 CT images, of which 740 were positive cases, and 325 were negative. They divided their dataset into training, test, and validation sets. Experiments showed that their model achieved 79.3\% accuracy, 83\% sensitivity, 67\% specificity, and 63\% F-measure in the test set.
	
	\section{Proposed Approach}
	\label{Proposed Approach}
	
	\subsection{Data sets and parameters}
	In this paper, the SARS-CoV-2 dataset is used to train and evaluate the model. SARS-CoV-2 is a publicly accessible database gathered by Angelov et al.
	\cite{soares2020sars}
	
	This dataset contains 2482 CT scan images collected from patients in Brazilian hospitals, including 1252 CT scan images belonging to COVID-19 cases. The rest of the CT scan images belong to healthy cases. Two hundred and fifty CT scan images from this dataset were randomly selected for the validation set. The same number of CT scan images were selected for the test set.
	
	It should be noted that the trained model is stored and used for the most accurate case on validation data. The batch size $b = 16$ was considered for the training phase. Adam was used as an optimizer for training the model. The learning rate was $lr = 0.003$.
	
	\begin{figure*}[]
		\centerline{\includegraphics[]{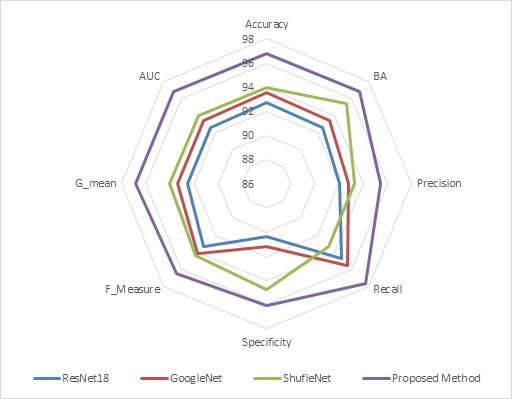}}
		\caption{Radar charts of the results of pre-trained networks and their combinations.}
		\label{fig2}
	\end{figure*}

	\begin{table*}[]
		\centering
		\caption{The results of pre-trained networks and their combinations.}
		\label{tab:1}
		\begin{tabular}{lllllllll}
			\hline
			Model           & Accuracy & BA     & Precision & Recall & Specificity & F-measure & G\_mean & AUC    \\ \hline
			ResNet18        & 92.742   & 92.563 & 92.029    & 94.776 & 90.351      & 93.382    & 92.537  & 92.563 \\ \hline
			GoogleNet       & 93.548   & 93.375 & 92..754   & 95.522 & 91.228      & 94.118    & 93.351  & 93.375 \\ \hline
			ShufleNet       & 93.952   & 95.42  & 93.284    & 93.284 & 94.737      & 94.34     & 94.007  & 94.01  \\ \hline
			Proposed Method & 96.774   & 96.843 & 95.402    & 97.647 & 96.04       & 96.512    & 96.84   & 96.843 \\ \hline
		\end{tabular}
	\end{table*}
	
	\subsection{Loss functions}
	Given that the purpose of the proposed model is to classify inputs in binary form, the model can be formulated as follows:
	\begin{equation}
		\hat{y}_{\theta, i}=\sigma\left(f_{\theta}\left(x_{i}\right)\right),
	\end{equation}
	where $f$ is the parameterized model with $\theta$. $\hat{y}$ is the predicted probability of the positive class. $\sigma$ is the sigmoid activation function defined as follows:
	\begin{equation}
		\sigma(z)=\frac{1}{1+\exp (-z)}.
	\end{equation}
	
	In general, the likelihood is as follows:
	\begin{equation}
		\label{eq:2}
		\mathbb{P}(\mathcal{D} \mid \theta)=\prod_{i=1}^{n} \hat{y}_{\theta, i}^{y_{i}}\left(1-\hat{y}_{\theta, i}\right)^{1-y_{i}},
	\end{equation}
	By taking the logarithm from Equation \ref{eq:2}, the following equation is obtained:
	\begin{equation}
		\log \mathbb{P}(\mathcal{D} \mid \theta)=\sum_{i=1}^{n}\left(y_{i} \log \hat{y}_{\theta, i}+\left(1-y_{i}\right) \log \left(1-\hat{y}_{\theta, i}\right)\right)
	\end{equation}
	Given that the goal of the classification model is in binary form, $y$ can have two values of 0 or 1. Therefore, for each $i$, either log of $(\hat{y}_{i})$ or log of $(1-\hat{y}_{i})$ is calculated. 
	
	$(\hat{y}_{i})$ and $(1-\hat{y}_{i})$ are the predicted probability that the data point $i$ is positive and the predicted probability that the data point $i$ is negative, respectively.
	
	Finally, because the logarithmic function is monotonic, maximization of the likelihood is equivalent to maximization of the logarithmic function (i.e., log-likelihood). To make things a little more complicated, because "minimizing loss" makes more sense, we can instead minimize the negative log-likelihood, reaching the well-known Negative Log-Likelihood Loss:
	 \begin{equation}
	 	l(\theta)=-\sum_{i=1}^{n}\left(y_{i} \log \hat{y}_{\theta, i}+\left(1-y_{i}\right) \log \left(1-\hat{y}_{\theta, i}\right)\right).
	 \end{equation}
	 
	\subsection{Backbones}
	Three pre-trained networks are used as the backbones of the proposed model. These networks are ResNet, GoogleNet, and ShuffleNet. The description of each of these networks is as follows:
	
	\textbf{ResNet}: It is an artificial neural network (ANN). A deep network with 152 layers can be made by ResNet. ResNet has introduced a skip connection to adapt the input from the previous level to the next without changing the input. Skipping the connection allows for a deeper network. The main reason to use a skip connection is to prevent the gradient from disappearing or exploding. Vanishing/exploding gradients occur during each iteration of training. Each of the neural network weights would be proportionally updated by the partial derivative of the loss function. Vanishing/exploding gradients effectively prevent the weight from changing its value in some cases. In the worst case, this can completely prevent the neural network from training further.

	\begin{figure*}
		\begin{subfigure}{.5\textwidth}
			\centering
			\includegraphics[width=.8\linewidth]{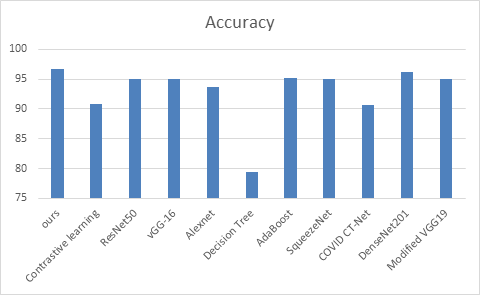}  
			\caption{
				Bar chart comparing the proposed method’s
				\\
				accuracy with other methods.}
			\label{fig3}
		\end{subfigure}
		\begin{subfigure}{.5\textwidth}
			\centering
			\includegraphics[width=.8\linewidth]{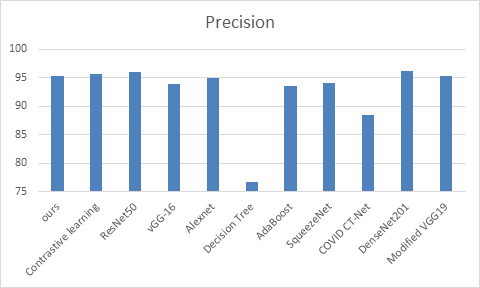}  
			\caption{Bar chart comparing the proposed method’s 
				\\
				precision with other methods.}
			\label{fig4}
		\end{subfigure}		
		\begin{subfigure}{.5\textwidth}
			\centering
			\includegraphics[width=.8\linewidth]{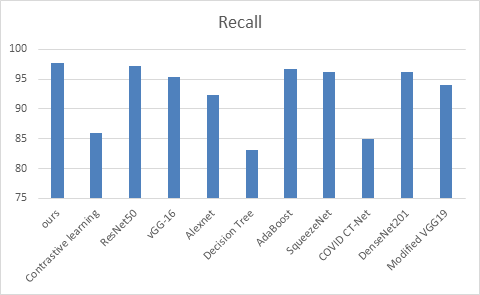}  
			\caption{Bar chart comparing the proposed method’s 
				\\
				recall with other methods.}
			\label{fig5}
		\end{subfigure}
		\begin{subfigure}{.5\textwidth}
			\centering
			\includegraphics[width=.8\linewidth]{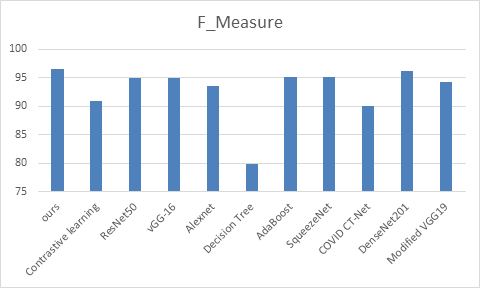}  
			\caption{Bar comparing the proposed method’s 
				\\
				F-measure with other methods.}
			\label{fig6}
		\end{subfigure}
		\caption{Bar charts for comparing different methods}
		\label{fig:fig}
	\end{figure*}
	
	\begin{table*}[]
		\centering
		\caption{Comparing the accuracy, precision, recall and f-measure of the proposed method with other methods.}
		\label{tab:2}
		\begin{tabular}{lcccc}
			\hline
			& \textbf{Accuracy} & \textbf{Precision} & \textbf{Recall} & \textbf{F-measure} \\ \hline
			\textbf{ours}                 & \textbf{96.774}   & 95.402             & \textbf{97.647} & \textbf{96.512}    \\ \hline
			\textbf{Contrastive learning} & 90.83             & 95.75              & 85.89           & 90.87              \\ \hline
			\textbf{ResNet50}             & 94.96             & 96                 & 97.15           & 95.03              \\ \hline
			\textbf{vGG-16}               & 94.96             & 94.02              & 95.43           & 94.97              \\ \hline
			\textbf{Alexnet}              & 93.75             & 94.98              & 92.28           & 93.61              \\ \hline
			\textbf{Decision Tree}        & 79.44             & 76.81              & 83.13           & 79.84              \\ \hline
			\textbf{AdaBoost}             & 95.16             & 93.63              & 96.71           & 95.14              \\ \hline
			\textbf{SqueezeNet}           & 95.1              & 94.2               & 96.2            & 95.2               \\ \hline
			\textbf{COVID CT-Net}         & 90.7              & 88.5               & 85              & 90                 \\ \hline
			\textbf{DenseNet201}          & 96.2              & \textbf{96.2}      & 96.2            & 96.2               \\ \hline
			\textbf{Modified VGG19}       & 95                & 95.3               & 94              & 94.3               \\ \hline
		\end{tabular}
	\end{table*}
	
	\textbf{GoogleNet}: This network won the ImageNet large-scale visual recognition competition (ILSVRC) in 2014. To reduce computation, GoogleNet uses 1$\times$1 convolution as a dimensionality reduction engine. 1$\times$1 convolution is a vector of size $f_{1}$ that convolves the entire image and produces an m×n output filter. If the number of 1$\times$1 convolutions is $f_{2}$, the output of all 1$\times$1 convolutions will be the size $(m, n, f_{2})$. Therefore, $f_{2}<f_{1}$ can be considered to represent an $f_{1}$ filter rather than an $f_{2}$ filter. It lets the network train how to reduce the dimension most efficiently.
	
	\textbf{ShuffleNet}: It is a CNN architecture that performs highly efficiently in computation. This architecture was designed for mobile devices. It has a computing power of 10–150 mega floating-point operations per second (MFLOPs).
	
	ShuffleNet uses point-by-point group convolution and shuffled channels. It causes a reduction in computational costs while maintaining accuracy. Compared with MobileNet systems, ShuffleNet achieves lower Top1 errors in classification on ImageNet. Compared with AlexNet, it also achieves a speed increase of about 13 times.
	
	\subsection{Model}
	
	The power of pre-trained networks in image classification has been well demonstrated. The question is, can the power of these networks be used together? In this study, pre-trained networks are placed side by side in parallel, and their feature vectors are concatenated. The outputs of the ResNet, GoogleNet, and ShuffleNet networks are vectors with a size of 1$\times$1000. 
	
		\begin{figure*}[]
		\centerline{\includegraphics[width=0.4\textwidth]{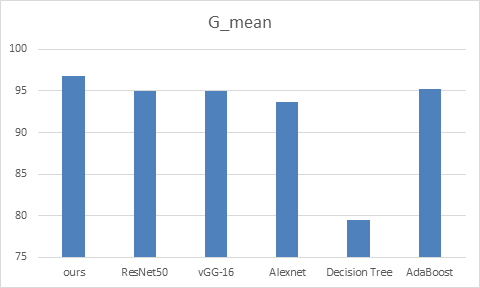}}
		\caption{Bar chart to compare the G\_mean of the proposed method with other methods.}
		\label{fig7}
	\end{figure*}
	
	\begin{table*}[]
		\centering
		\caption{Comparing the G\_mean of the proposed method with other methods.}
		\label{tab:3}
		\begin{tabular}{cccccccc}
			\hline
			& \textbf{ours} & \textbf{ResNet50} & \textbf{VGG-16}
			& \textbf{Alexnet} & \textbf{Decision Tree} & \textbf{AdaBoost} 
			\\ \hline
			\textbf{G\_mean} & 96.84         & 94.98             & 94.96           & 93.68               & 79.51              & 95.19             \\ \hline
		\end{tabular}
	\end{table*}
	
	These vectors are placed side by side to form a 3,000-neuron layer. The 3000-neuron layer is fully connected to a 512-neuron layer. The 512-neuron layer is fully-connected to a 10-neuron layer. Using the resulting vector with size 10, the loss value is calculated using the negative log-likelihood relations.
	The proposed network is trained using the calculated loss and "Back Propagation" algorithm. The Adam optimizer was used for training. It should be noted that the images are fed to three backbones simultaneously for training. Model training is such that during 50 epochs, the epoch model with the highest accuracy is selected for the testing phase. Figure
	\ref{fig1}
	shows the architecture of the proposed method. The proposed model will have about 22 million parameters. This model is trained at a rate of 0.2 for dropout.
			
	\section{Experiments}
	\label{Experiments}
	
	\subsection{Evaluation measures}
	\label{Evaluation measures}
	This section describes the evaluation criteria for the results of the best-trained model. Eight evaluation criteria were used to evaluate the results. These criteria are accuracy, precision, recall, F\_measure, specificity, balanced accuracy (BA), geometric mean (G-mean), and Area under the curve (AUC). These criteria are calculated as follows:
	\begin{equation}
		\text { Accuracy }=\frac{T P+T N}{T P+T N+F P+F N},
	\end{equation}
	\begin{equation}
		\label{eq:Precision}
		\text { Precision }=\frac{T P}{T P+F P},
	\end{equation}
	\begin{equation}
		\label{eq:Recall}
		\text { Recall }=\frac{T P}{T P+F N},
	\end{equation}
	\begin{equation}
		\label{eq:fmeasure}
		F_{-}{\text {measure }}=\frac{2 \times \text { Precision } \times \text { Recall }}{\text { Precision }+\text { Recal }},
	\end{equation}
	\begin{equation}
		\text { Specificity }=\frac{T N}{T N+F P},
	\end{equation}
	\begin{equation}
		B A=\frac{\text { Recall }+\text { Specificity }}{2},
	\end{equation}
	\begin{equation}
		\label{eq:Gmean}
		G_{-} \text {mean }=\sqrt{\text { Specificity } \times \text { Recall }},
	\end{equation}
	where $TP$ is true-positive, $TN$ is true negative, $FP$ is false-positive, and $FN$ is false-negative in the confusion matrix.
	The area under the "recall in terms of $FPR$" curve should be calculated to calculate the AUC. $FPR$ is calculated as follows:
	\begin{equation}
		F P R=\frac{F P}{T N+F P}=1-\text { Specificity }.
	\end{equation}
	\subsection{Experimental results}
	The results obtained from the proposed model on the test data are calculated by the criteria introduced in Section \ref{Evaluation measures}. The results of the proposed model are presented in Table \ref{tab:1}. It should be noted that using the negative log-likelihood loss, each of the pre-trained networks ResNet, GoogleNet, and ShuffleNet were fine-tuned separately with the SARS-CoV-2 dataset. The fine-tuning network process is similar to the training procedure in the proposed model. In addition to the results of the proposed model, the results of each of the ResNet, GoogleNet, and ShuffleNet networks on the test data are presented in Table \ref{tab:1}.
	As shown in Table \ref{tab:1}, the proposed model performs better than each of the models separately.
	
	Figure \ref{fig2} is based on the data in Table \ref{tab:1}. As shown in Figure \ref{fig2}, the performance of the combined model is better than each of the individual models.
	
	Table \ref{tab:2} shows the results of four evaluation criteria including accuracy, precision, recall, and F-measure for the proposed method and ten other methods. These ten methods are Contrastive learning
	\cite{wang2020contrastive},
	ResNet50
	\cite{ai2020correlation},
	vGG-16
	\cite{aria2022ada},
	Alexnet
	\cite{soares2020sars},
	Decision Tree
	\cite{bernheim2020chest},
	AdaBoost
	\cite{zhao2020relation},
	SqueezeNet
	\cite{aria2022ada},
	COVID CT-Net
	\cite{yazdani2020covid},
	DenseNet201
	\cite{jaiswal2021classification},
	and Modified VGG19
	\cite{panwar2020deep}.
	
	Figure \ref{fig3} shows a bar chart comparing the proposed method’s accuracy with that of ten other methods. The accuracy obtained for the proposed method (96.774\%) is the highest Accuracy value among other methods. After the proposed method, DenseNet201 has the highest accuracy value with 96.2\%. Among the proposed methods, Decision Tree has the lowest accuracy with 79.44\%.
	
	Figure \ref{fig4} shows a bar chart comparing the proposed method’s precision with ten other methods. The precision of the proposed method (95.402\%) along with DenseNet201, ResNet50, and Contrastive learning methods is the highest precision value. According to Equation \ref{eq:Precisio}, precision is equal to the fraction of the number of items correctly detected by the model over the number of items that are actually true. Based on this, precision can be considered a measure of the model's predictive power. According to Figure \ref{fig4}, the proposed method is one of four powerful methods for prediction.
	
	Figure \ref{fig5} shows a bar chart comparing the proposed method’s recall with that of ten other methods. According to Equation \ref{eq:Recall}, recall refers to the percentage of total predictions that the model correctly categorizes. Among the presented methods, the proposed method has the highest recall value (97.647\%). The Contrastive learning, Decision Tree, and COVID CT-Net methods have the lowest recall value.
	
	Figure \ref{fig6} shows a bar chart comparing the proposed method’s F-measure with that of ten other methods. The proposed method and DenseNet201, with values above 96\%, have the highest value for F-measure. F-measure is calculated using Equation \ref{eq:fmeasure}. In fact, F-measure is a type of average between precision and recall used in statistics to evaluate the performance of systems.
	
	Table \ref{tab:3} and Figure \ref{fig7} show the results of the G-mean calculation based on Equation \ref{eq:Gmean} for the proposed method and the five methods ResNet50, VGG-16, Alexnet, Decision Tree, and AdaBoost. The proposed method has the highest G-mean value at 96.84\%.
		
	\section{Conclusion}
	\label{Conclusion}
	In this paper, a combination of three well-known pre-trained networks for the diagnosis of COVID-19 disease is presented. According to the loss used, it can be concluded that on some datasets such as SARS-CoV-2, certain combinations of different architectures can improve the detection performance.
	By combining three pre-trained networks in parallel, the accuracy increased by about 3\%. In addition to the increase in accuracy, other evaluation criteria were also improved.


\end{document}